\documentstyle[emulateapj,onecolfloat,psfig]{article}

\newcommand{\lin}{{\rm lin}}
\newcommand{\nlin}{{\rm nl}}

\newcommand{\fsky}{f_{\rm sky}}

\newlength{\tskip}
\newlength{\colwidth}

\newcommand{\beq}{\begin{equation}}
\newcommand{\eeq}{\end{equation}}
\newcommand{\beqa}{\begin{eqnarray}}
\newcommand{\eeqa}{\end{eqnarray}}

\begin{document}
\twocolumn[
\title{Measuring Angular Diameter Distances through Halo Clustering}
\author{Asantha Cooray$^{1}$\altaffilmark{2}, Wayne Hu$^1$, Dragan Huterer$^3$, and Michael Joffre$^1$}
\affil{$^1$Department of Astronomy and Astrophysics, University of Chicago,
Chicago, IL 60637\\
$^3$Department of Physics, University of Chicago, Chicago, IL 60637\\
E-mail: (asante,whu,dhuterer,joffre)@oddjob.uchicago.edu}
\altaffiltext{2}{Sherman Fairchild Senior Research Fellow, Theoretical
Astrophsyics, California Institute of Technology, Pasadena, CA 91125}

\begin{abstract}
Current and upcoming wide-field surveys for weak gravitational lensing 
and the Sunyaev-Zel'dovich
effect will generate mass-selected catalogues of dark matter halos
with internal or followup photometric redshift information. 
Using the shape of the linear power spectrum as a standard ruler that
is calibrated by CMB measurements, we find
that a survey of 4000 deg.$^2$ and a mass threshold of $10^{14}
M_{\sun}$ can be used to determine the comoving angular diameter
distance as a function of redshift.  In principle, this test also allows an
absolute calibration of the distance scale and measurement of the
Hubble constant.  This test is largely insensitive
to the details of halo mass measurements, mass function, and halo bias. 
Determination of these quantities would further allow a measurement of the
linear growth rate of fluctuations.  
\end{abstract}


]
\section{Introduction}

A number of observational efforts are now underway or being planned 
to image the large-scale structure of the universe spanning a range of 
redshifts.   These wide-field surveys typically cover tens to
thousands of square degrees of the sky: the ongoing Sloan Digital
Sky Survey
(SDSS), the weak
gravitational lensing shear observations with instruments such as
SNAP or
the Large Aperture Synoptic Survey Telescope,
and surveys of the Sunyaev-Zel'dovich effect (SZ;
\cite{SunZel80} 1980).

In addition to their primary science goals, these surveys are
expected to produce catalogues of dark matter halos, which in the
case of lensing and SZ surveys are expected to be essentially
mass selected (\cite{Witetal01} 2001; \cite{Holetal00} 2000).
Lensing and other optical surveys are particularly promising in
that they will provide photometric redshifts on the member
galaxies of a given halo (e.g., \cite{Hog98} 1998); 
this will render accurate determination of
the halo redshift.  Halo number counts as a function of redshift
is a well-known and powerful cosmological test. 
Here we consider the additional
information supplied by the angular clustering of halos.

A feature in an angular power spectrum of
known physical scale and originating from a known redshift
can be used to measure the angular diameter
distance between us and this redshift; this has most notably
been applied to the case of the cosmic microwave background (CMB) to
determine the distance to redshift $z \sim 10^3$. The angular
power spectrum of halos provides a similar test based on the 
standard ruler defined by its shape.  
In the adiabatic cold dark matter model for structure formation,
this standard ruler is essentially the horizon at matter-radiation
equality and its absolute physical scale can be directly calibrated with
CMB anisotropy data.  In principle then, one can determine
the angular diameter distance as a function of redshift and test the properties
of the dark energy.

As a purely geometric test, this method is largely insensitive
to uncertainties in the the mass function and the relationship between
the halo masses and the actual observables, e.g.\ the SZ temperature decrement
or lensing aperture mass.   The bias of the halos
is scale-dependent only on small (non-linear) scales,
and can in principle be extracted to arbitrary precision from
$N$-body simulations.  If and when these quantities are securely 
known, one can extract further information from the amplitude 
and small-scale behavior of the power spectrum.  In particular,
the linear growth rate and non-linear scale provide extra handles
on the dark energy.

For illustrative purposes, we adopt the $\Lambda$CDM cosmology with 
energy densities (relative to critical) of $\Omega_m=0.35$ in matter,
$\Omega_b=0.05$ in baryons, $\Omega_\Lambda=0.65$ in vacuum
energy, the dimensionless Hubble constant of $h=0.65$, and a
scale-invariant spectrum of primordial fluctuations, normalized
to the present day galaxy cluster abundance ($\sigma_8=0.9$;
\cite{ViaLid99} 1999).  

%

\begin{figure}[t]
\centerline{\psfig{file=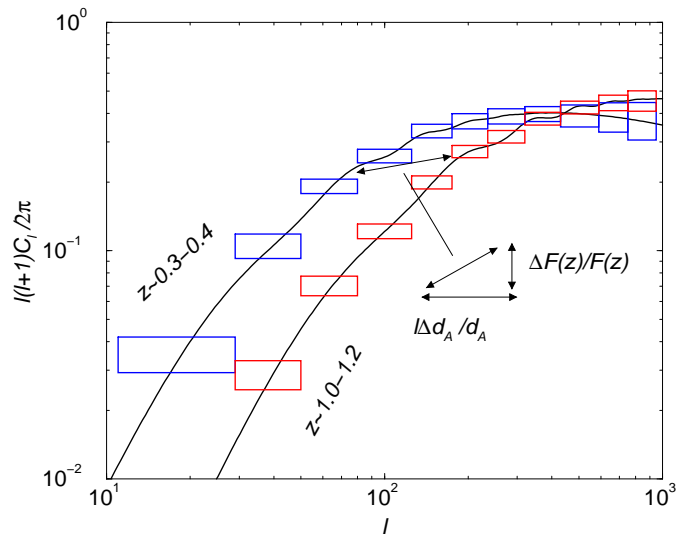,width=3.5in,angle=-90}}
\caption{Angular power spectrum of halos with $M> 10^{14} M_{\sun}$ 
in a wide-field survey
in bins of $z=0.3-0.4$ and $1.0-1.2$.  The binned
errors are 1-$\sigma$, and assume a survey of 4000 deg$^2$, within reach
of upcoming weak lensing and SZ surveys. The angular
power spectrum at high redshifts is shifted towards the right
proportional to the increase in the comoving angular diameter distance. 
The oscillations are due to baryons but we ignore the additional information
they contain.}
\label{fig:halo}
\end{figure}

\section{Angular Power Spectrum}
\label{sec:fisher}

The angular power spectrum of halos in $i$th redshift bin 
is a Limber (1954; \cite{Kai92} 1992) projection of the 
the halo number density power spectrum 
\begin{equation}
C_l^i = \int dz\,
W^2_i(z) {H(z) \over d_A^2(z)} P_{hh}\left(\frac{l}{d_A}; z\right) \,,
\label{eqn:cl}
\end{equation}
where $W_i(z)$ is the distribution of halos in a given redshift bin normalized
so that $\int\, dz\, W_i(z)=1$,
$H(z)$ is the Hubble parameter, and $d_A$ is the
angular diameter distance in comoving coordinates.
Note that $W_i(z)$ comes directly from the observations of the number
counts as a function of redshift and depends on the mass function and 
mass sensitivity of the employed observable.

If the halos trace the linear density field, 
\begin{equation}
P_{hh}(k;z) = \left< b_M \right>^2(z) D^2(z) P^\lin(k;0)\,,
\end{equation}
where $\left<b_M\right>$ is the mass-averaged halo bias
parameter, $P^\lin(k;0)$ is
the present day matter power spectrum computed in linear theory, and $D(z)$
is the linear growth function
$\delta^\lin(k;z) = D(z)\delta^\lin(k;0)$.  A scale-independent halo bias
is commonly assumed in the so-called ``halo model'' (e.g. \cite{Sel00} 2000) 
and should be valid at least in the linear regime.
Equation~(\ref{eqn:cl}) then becomes
\begin{eqnarray}
C_l^i & = & \int dz
 \,W_i^2(z) F(z) P^\lin\left(\frac{l}{d_A^i};0\right) \,, \\
\label{eqn:cz}
F(z) & = & {H(z) \over d_A^2(z)}  D(z)^2 \left< b_M \right>^2(z)\,.
\label{eqn:fz}
\end{eqnarray}

The underlying linear power spectrum contains two physical scales:
the horizon at matter
radiation equality
\begin{equation}
k_{\rm eq} = \sqrt{2 \Omega_m H_0^2 (1+z_{\rm eq})} \propto \Omega_m h^2
\end{equation}
which controls the overall shape of the power spectrum, and the sound 
horizon at the end of the Compton drag epoch, $k_{\rm s}(\Omega_m
h^2,\Omega_b h^2)$, which
controls the small wiggles in the power spectrum.
The angular or multipole locations of these features shift in
redshift as $l_{\rm eq, s} = k_{\rm eq, s} d_A(z_i)$.
We propose the following test: measure
$C_l^i$ in several redshift bins and, using the fact that 
 $l_{\rm eq}$  scales with $d_A(z_i)$, 
constrain the  angular diameter distance as a function of redshift.
To be conservative, we ignore the additional information suppled by
$l_{\rm s}$.

In Fig.~\ref{fig:halo}, we illustrate the proposed test.  
The two curves show the halo power spectra in two redshift bins:
$0.3<z<0.4$ and $1.0<z< 1.2$. The angular power spectrum
corresponding to the higher redshift bin is shifted to the right
in accordance to the ratio of angular diameter distances $(\delta l/l
\sim \delta d_A/d_A)$. Much of this shift simply reflects the
Hubble law, $d_A \approx z/H_0$.  Since the physical scale of the 
two features --- the overall shape of the
spectrum and the baryon oscillations --- can be calibrated
from the morphology of the CMB peaks, these measurements can in principle 
be used to determine the Hubble constant independently of the 
distance ladder and distance to last scattering surface.

In addition to the horizontal shift due to the change in angular diameter
distance, the power spectra in Fig.~\ref{fig:halo} are shifted
vertically due to the change in $F(z)$ (Eq.~\ref{eqn:fz}).
By ignoring the information contained in $F(z)$, this purely geometric test 
is robust against uncertainties in the mass selection, mass 
function and linear bias.  Of course, if these uncertainties 
are pinned down independently, both $F(z)$ and the halo abundance
in $W_i(z)$ will help measure the growth rate of structure.

\section{Parameter Estimation}


Even though the angular diameter distance test is robust against uncertainties
in the halo selection function, number density and bias, these quantities 
enter into the consideration of the signal-to-noise for a realistic
survey.  We will focus on a survey of 4000 deg.$^2$ with a detection
threshold in mass of $10^{14}$ M$_{\sun}$ out to $z=2$. We use
a total of 9 bins in redshift; note that the cluster photometric redshift
accuracy is expected to be much smaller than the bin width.
The mass threshold is consistent with those
expected from upcoming lensing and SZ effect surveys 
(see \cite{KruSch99} 1999;
\cite{Holetal00} 2000; Joffre et al.\ in preparation) and the survey
area is consistent with a planned SZ survey from the South Pole
Telescope (Carlstrom, private communication).
To compute $W_i^2(z)$ we adopt the predictions of the Press-Schechter 
mass function (PS; \cite{PreSch74} 1974).  This mass function, along with
the halo bias prescription of \cite{MoWhi96} (1996), is also used
to predict the mass-averaged halo bias $\left<b_M\right>(z)$.

Assuming Gaussian statistics, we can express the uncertainty in the 
measurements of the angular power spectrum as
\begin{equation}
\Delta C_l^i = {\left(C_l^i + N_l^i\right) \over \sqrt{(l+1/2)\fsky}},
\label{eqn:delta}
\end{equation}
where $\fsky = \Theta_{\rm deg}^2 \pi/129600 $ is the fraction of
the sky covered by a survey of dimension $\Theta_{\rm deg}$ in
degrees and $N_l^i$ is the noise power spectrum. We assume that
the dominant source of noise is the shot-noise so that
$N_l^i \equiv 1/\bar{N}_i$, where $\bar{N}_i$ is the surface density
of the halos in the $i$th redshift bin.   We 
use the PS mass function to predict $\bar{N}_i$.
In Fig.~\ref{fig:halo}, the two bins contain roughly $\sim$ 4 and 6
halos/deg.$^2$ above our minimum mass.
In the same Figure, we show band power measurement errors
following Eq.~\ref{eqn:delta}.

\begin{figure}[t]
\centerline{\psfig{file=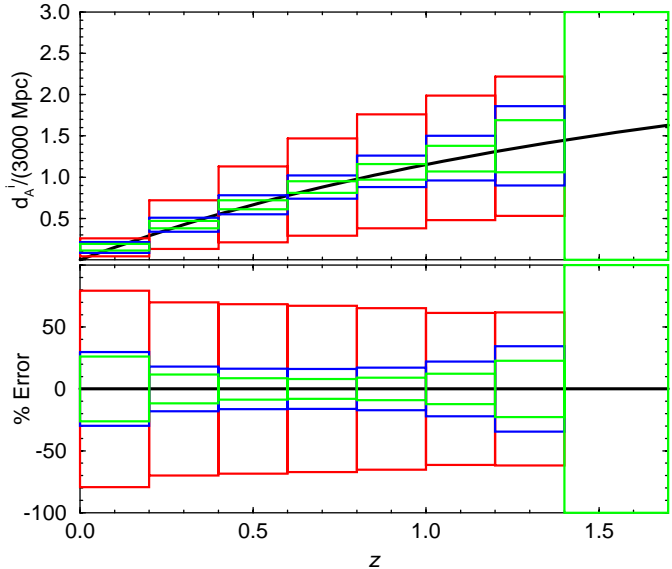,width=3.5in,angle=0}}
\caption{(a) The errors (1-$\sigma$) on angular diameter distance
as a function of redshift.  We have binned the halos in 8
redshift bins between 0 and 1.6. The larger errors are with no
prior assumption on the cosmological parameters that define the
transfer function while the smaller errors are with MAP (Temp) and
Planck (Pol) priors. In (b), we show relative errors in the distance.}
\label{fig:da}
\end{figure}

\label{sec:model}

To estimate how well halo clustering can recover cosmological
information, we construct the Fisher matrix
\begin{equation}
{\bf F}_{\alpha \beta} = \sum_{i=1}^{{\rm N_{\rm bins}}} 
\sum_{l=l_{\rm min}}^{l_{\rm max}^i}
	{(l + 1/2) \fsky \over (C_l^i + N_l^i)^2} 
	{\partial C_l^i \over \partial p_\alpha}
	{\partial C_l^i \over \partial p_\beta}\,,
\label{eqn:Fisher}
\end{equation}
where $\alpha$ and $\beta$ label parameters that underly the power spectra.
Since the variance of an unbiased estimator of a
parameter $p_\alpha$ cannot be less than $({\bf F}^{-1})_{\alpha
\alpha}$, the Fisher matrix quantifies the best statistical
errors on parameters possible with a given data set.
 
\begin{figure}[b]
\centerline{\psfig{file=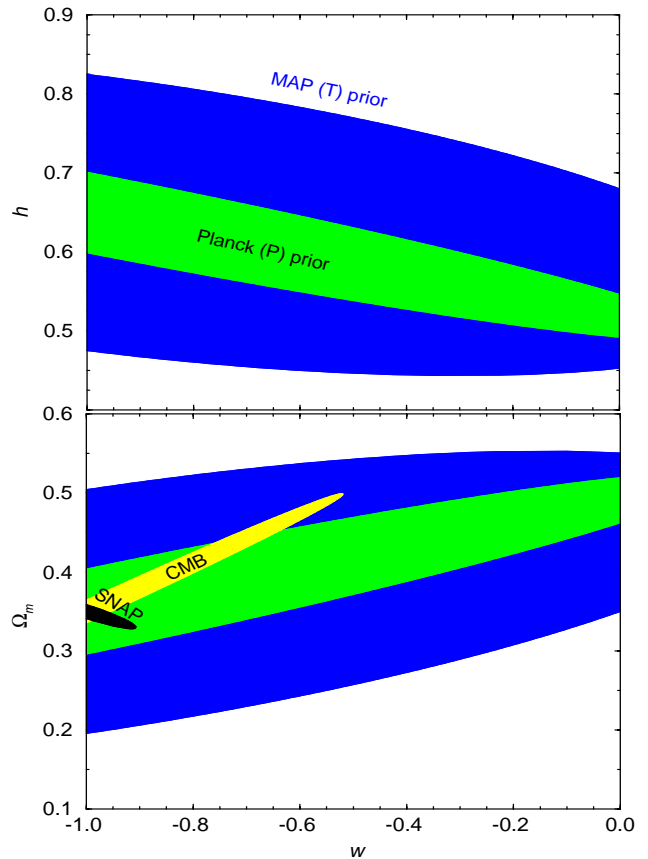,height=4.5in,width=3.3in,angle=0}}
\caption{The errors (all 1-$\sigma$) on (a) $h$ and $w$ and (b)
$\Omega_m$ and $w$,  using distance information only. In both (a) and (b)
we show errors for halos with priors following MAP (Temp) and
Planck (Pol). In (b), for comparison, 
we also show errors on $\Omega_m$ and $w$ from CMB 
(Planck with temperature and polarization (\cite{HuEisTegWhi99} 1999)) 
and type Ia SNe with SNAP mission.}
\label{fig:w1}
\end{figure}

We choose $l_{\rm min}=2\pi/\Theta_{\rm deg}$ when evaluating
equation~(\ref{eqn:Fisher}) as it corresponds roughly to the
survey size.  The precise value does not matter for parameter
estimation due to the increase in sample variance on the survey
scale.  Given our crude Gaussian approximation of the shot-noise,
we choose a conservative $l_{\rm max}^i$ corresponding to the
multipole at which the noise and sample variances are equal
$N_{l}^i = C_{l}^i$: $l_{\rm max}^i$ ranges from 200 at low
redshift bins to 400 at high redshift.  At low redshifts
this cutoff is slightly in the non-linear regime, but at
redshifts greater than 0.8 or so one is well within the linear
regime. Therefore, such a low $l_{\rm max}$ largely eliminates uncertainties
in the modeling of scale-dependent halo bias in the non-linear regime.

\begin{figure}[b]
\centerline{\psfig{file=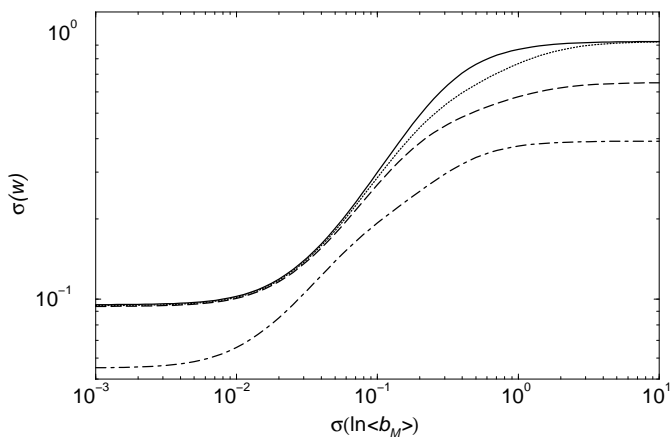,width=3.5in,angle=-90}}
\caption{The 1-$\sigma$ error on $w$ as a function of the prior
on bias. The four curves assume
Planck (Pol) priors on $\Omega_mh^2$, $n_s$ and $\Omega_bh^2$ to
define the linear power spectrum. The solid line is with no prior on
$\ln A$, and $h$. The dotted line includes a prior of 0.2 in $\ln A$,
while the long dashed line is with an addition prior of 0.1 in $h$.
The dot-dashed line is a highly optimistic scenario with  exact
$A$, a prior of 0.1 in $h$, and  using halo angular power spectrum 
information out to $l_{\rm max}$ of 1000, deeply within the non-linear
regime.}
\label{fig:wg}
\end{figure}

Due to the dependence of $C_l^i$ on $P^\lin(k)$, all cosmological
parameters that change the shape of the matter power spectrum
across the scales probed by halos also affect the measurement of
distance. The shape of the transfer function is determined by
$\Omega_mh^2$ and $\Omega_bh^2$ while the overall slope is
determined by a scalar tilt $n_s$.  
Note that these parameters
will be accurately determined from CMB anisotropy observations.
Since the CMB peaks probe the
same range in spatial scale as the halo power spectrum, one
is  relatively insensitive to deviations from a pure initial power law.

When estimating expected errors on distance, we will consider
several sets of priors on these cosmological parameters. These
priors follow Table~2 of \cite{Eisetal99} (1999) and correspond
to constraints expected from MAP and Planck with and without
polarization.  Though baryon oscillations contain cosmological
information,  in order to be conservative 
against possible non-linearities in the bias we ignore the 
information present in the baryon oscillations, and employ the
smooth fitting function of \cite{EisHu99} (1999).  Our results
are then very weakly dependent on the fiducial value or priors 
on $\Omega_b h^2$. In case baryonic features in the
angular power spectrum are detected, we expect additional cosmological
information to be gained using the proposed test.

\section{Results \& Discussion}

We first consider the measurement of the angular diameter distance
$d^i_{A}= d_A(z_i)$.  In addition to the cosmological parameters
$\Omega_m h^2$, $\Omega_b h^2$ and $n_s$, 
we include a set of parameters $F^i=F(z_i)$ which allow the normalization
of the $i$ power spectra to float independently.  Both $d^i_{A}$
and $F^i$ approximate underlying functions as piecewise flat
across each bin.

In Fig.~\ref{fig:da}(a), we show three sets of errors: the largest
errors assume no prior knowledge on the transfer function, while
the smaller errors correspond to priors from MAP (Temp) and Planck
(Pol) respectively (\cite{HuEisTegWhi99} 1999). 
In Fig.~\ref{fig:da}(b), we show the
fractional percentage errors on the distance. The errors in the lowest bins
tell us how well one can estimate the Hubble constant, while
the slope of $d_A(z)$ around $z \sim 1$ provides information on
cosmology.  $d_A$ is best determined around $z \approx 0.7$ at the
level of 10\%. 
For systematic errors in the redshifts to dominate the
error budget, the inferred mean redshift of the bin must differ from the
true value by $\sim 10\%$,
which is far above the errors we expect for individual halos even from 
photometric techniques at $z < 1$.  Precision at higher redshift is
not required because of the large statistical errors.
The errors on distance estimates $d_A^i$
are correlated at the 5\% level due to remaining 
uncertainties in parameters that affect all $d_A^i$
(e.g., $\Omega_m h^2$).

In a realistic cosmology $d_A(z)$ is smoothly varying.
 Since $\Omega_m h^2$ is already taken as
a parameter, we parameterize $d_A$ with the Hubble constant
$h=H_0/100$km s$^{-1}$ Mpc$^{-1}$ and the equation of 
state of the dark energy $w$, the ratio of pressure to density, 
assuming a flat Universe.  
We bin halos
following the binning scheme in Fig.~\ref{fig:da}. 
In Fig.~\ref{fig:w1}a we show that a strong degeneracy
in $h$ and $w$ 
remains even with Planck priors because $d_A$ is only accurately recovered in a small
redshift range.  Of course, an external determination of $h$ would break
this degeneracy.

An alternate way of breaking the degeneracy is to employ other
cosmological probes of $\Omega_m$ and $w$.
As shown in Fig.~\ref{fig:w1}(b), different linear
combinations of $\Omega_m$ and $w$ will be determined by halos and
the CMB due to the different redshift ranges probed. 
Although each constraint alone may not be able to pin down
$w$, halos and the CMB combined allow very interesting
constraints even under our conservative assumptions. 

Uncertainties in the 
mass threshold and the scale dependence of halo bias 
are potential caveats to these conclusions.   A mass threshold that
differs from the assumed $10^{14}$ M$_{\sun}$ value would not bias
the angular diameter distance results since they only utilize redshift and
power spectrum shape information. However it would affect the errors
due to a rapid decrease in the number density of halos with threshold mass:
at $4 \times 10^{14}$ M$_{\sun}$ the error on $w$ increases 
increase by a factor of $\sim$ 3. 

A scale-dependent bias that can be {\it predicted} actually
{\it aids} in the determination of angular diameter distances: the scale-dependence
acts as another standardizable ruler for the test.  Indeed, 
the scale-dependence of the bias as a function of halo mass
is something that can be precisely
determined from $N$-body simulations (\cite{KraKly99} 1999). 
A more subtle problem is introduced by the addition of uncertainties in 
the mass threshold or selection function.  
Since the bias is also mass-dependent,
the uncertainty in the mass threshold $\delta M/M=0.1$ translates into the
uncertainty in the mass-averaged bias $\delta \left< b_M \right>/\left<
b_M \right>=0.03$. 
To investigate a scale dependent bias, we model it as 
\begin{equation}
b_M(k,z) = \left< b_M \right>(z) \left[1 + f \left(
\sqrt{\frac{P^\nlin(k;z)}{P^\lin(k;z)}} - 1\right)\right] \, .
\label{eqn:scalebias}
\end{equation}
where $f$ is a dimensionless parameter meant to interpolate bias between
the linear ($f \rightarrow 0$) and the non-linear ($f \rightarrow 1$)
regimes. Note that in the halo approach to clustering, 
 the non-linear mass power spectrum is a sum of the halo power spectrum
and contributions due to dark matter within halos; therefore, the halo
power spectrum cannot be larger than the non-linear power spectrum.
Taking a fiducial model with $f=0$ and adopting MAP (Temp) priors, 
we find that marginalizing over $f$ increases the error on $w$ by
less than ten percent.

To the extent that the mass threshold, halo bias, and mass function are
known, the amplitude of the halo power spectra can be
used to measure the linear growth rate. 
We conclude by estimating the level at which these quantities must
be determined to yield additional constraints on $w$.
In Fig.~\ref{fig:wg}, we 
plot the marginalized errors on $w$ as a function of the
assumed fractional prior on $\left<b_M^i \right>$
for various independent constraints on
$A$ and $h$. Since cosmological information captured 
in linear growth is determined by
relative amplitude variations in $C_l^i$,
the knowledge of the overall normalization $A$ is not crucial.
 For example, going from no prior knowledge of $A$ to a Gaussian prior with width of
20\% of the fiducial value of $A$
results in a decrease in $\sigma(w)$ of $\sim$ 25\%.
Errors on the mass selection function bias the measure of $w$. 
Using an extension to the Fisher
matrix approach, we determined that a 25\% systematic offset in mass
threshold from the fiducial value of 10$^{14}$ M$_{\sun}$
leads to systematic
bias in $w$ of 0.05 from its fiducial value of -1.0.  
Lensing simulations 
(\cite{MetWhi99} 1999; \cite{RebBar99} 1999) indicate
that the calibration of projection effects at this level will 
be challenging but feasible to achieve. 

Clearly future surveys which can identify 
dark matter halos as a function of redshift contain valuable information
beyond the evolution of their number abundance.  As the theoretical 
modeling of the halo distribution and empirical modeling of the
selection process improve, the correlation function of the halos
can provide not only the angular diameter distance, 
but also direct measurements of the growth of large-scale structure. 

{\it Acknowledgements:} A.C. and W.H were supported by NASA NAG5-10840
and DOE OJI. D.H. was supported by the DOE.

\end{document}